# Engineering and improving the magnetic properties of thin Fe layers through exchange coupling with hard magnetic Dysprosium layers

M. Ehlert, H. S. Körner, T. Hupfauer, M. Schitko, G. Bayreuther, and D. Weiss*

*Institute of Experimental and Applied Physics, University of Regensburg, D-93040 Regensburg, Germany*

We report on a comprehensive study of the magnetic coupling between soft magnetic Fe layers and hard magnetic Dysprosium (Dy) layers at low temperatures (4.2 - 120 K). For our experiments we prepared thin films of Fe and Dy and multilayers of Fe/Dy by ultra-high vacuum sputtering. The magnetic properties of each material were determined with a super conducting quantum interference device. Furthermore, we performed magnetoresistance measurements with similarly grown, microstructured devices, where the anisotropic magnetoresistance (AMR) effect was used to identify the magnetization state of the samples. By analyzing and comparing the corresponding data of Fe and Dy, we show that the presence of a Dy layer on top of the Fe layer significantly influences its magnetic properties and makes it magnetically harder. We perform a systematic evaluation of this effect and its dependence on temperature and on the thickness of the soft magnetic layer. All experimental results can consistently be explained with exchange coupling at the interface between the Fe and the Dy layer. Our experiments also yield a negative sign of the AMR effect of thin Dy films, and an increase of the Dy films' Curie temperature, which is due to growth conditions.

## I. INTRODUCTION

Semiconductor spintronic devices often engage ferromagnetic (FM) electrodes on top of a semiconductor channel [1-3]. By using the FM electrodes as spin-sensitive probes, a non-equilibrium spin accumulation can be generated and detected in the semiconductor channel. This becomes especially challenging for measurement setups where the magnetization **M** of the FM detector is required to stay aligned along a pre-defined axis, while an external magnetic field **B** is applied transverse to the magnetization axis. For instance, this approach is used in a geometry for the detection of the spin Hall effect (SHE) [4-6]. In these experiments the spin probes are initially magnetized parallel to their axis. After that, a magnetic field, which is transverse to the FM magnetization axis and lies in the plane of the sample, is applied. This induces a Hanle spin precession of the SHE generated spin accumulation which can now be detected with the FM electrodes. The FM magnetization, however, stays aligned along the pre-defined direction only for small values of the transverse field. For larger magnetic fields, it reorients parallel to the **B**-field, leading to a rapid decay of the SHE induced signal. To prevent such signal decay, it is crucial to tailor and to control the magnetic properties of the employed thin FM layer.

The goal of the present work is to improve the magnetic properties of soft FM Fe layers, which are commonly used as spin detector or injector [1, 2, 7], by a simple and widely applicable technique which makes use of exchange coupling between soft and hard magnetic materials. The first experimental observation of exchange (bias) coupling was reported for the ferromagnetic/antiferromagnetic (FM/AFM) bilayer Co/CoO by Meiklejohn and Bean in 1956 [8, 9]. In FM/AFM systems, the antiferromagnetic material acts as a pinning layer for the FM layer. This gives rise to an unidirectional magnetic anisotropy and a shift of the hysteresis loop of the FM, when the material is field cooled through the Néel temperature $T_N$ of the AFM (assuming that the Curie temperature $T_C$ of the FM is higher than $T_N$) [10]. Research in the past decades focused on AFM/FM interfaces, including multilayers of AFM/FM and interlayer exchange coupling through nonmagnetic spacer layers. Progress both in experimental realization as well as in theoretical description has been reviewed by several authors [10-12]. Comparable coupling effects can also be observed in bilayer and multilayer systems of soft and hard FMs [13-15] and even in bilayers of soft and hard ferrimagnets [16, 17]. In soft/hard FM systems, the hard FM material acts as a pinning layer for the soft FM. Based on the proposals for exchange-spring magnets by Kneller and Hawig [18] and Coey and Slomski [19], exchange-spring coupling has been experimentally observed and theoretically modelled for various hard/soft FM systems [20-23]. Here we explore the potential of Dy/Fe bilayers for hard magnetic contacs.

The paper is organized as follows. In Sec. II we describe the coupling mechanism between Fe and Dysprosium (Dy), the sample growth and the experimental techniques which we used in our studies. Data of Dy and Fe single layers is presented in Sec. III. and IV. Measurements of the magnetic interplay between both materials in a Fe/Dy bilayer are presented in Sec. V. The paper concludes with a summary in Sec. VI.

## II. EXCHANGE COUPLING IN FE/DY BILAYERS

For our experiments we use the rare-earth element Dy as the hard FM material to magnetically pin the soft FM Fe. Dy is known to have a large saturation magnetization up to several Tesla ($\mu_0 M_s = 3.75$ T) [24] and large coercive fields. Both features qualify Dy to enhance the magnetic stability of soft FM spin probes. Detailed studies on the magnetic properties of Dy [25-28] revealed that its magnetic properties are (like most rare-earth elements) different from those of the FMs of the iron group. With decreasing temperature, Dy undergoes different magnetic phase transitions [29-32]. Dy is paramagnetic at room-temperature, becomes antiferromagnetic with a helical phase below the Néel temperature $T_N = 180$ K, and finally it turns into an ordinary FM below the Curie temperature $T_C = 90$ K. Temperatures below 90 K are used in most spin injection and detection experiments.

In Fe/Dy bilayers, the coupling between the rare-earth material Dy and the transition metal Fe occurs through exchange coupling at the interface of both materials [33, 34]. The magnetic properties of Dy in its FM phase mainly derive from its 4f electrons, whereas those of Fe incorporate its 3d electrons. Since the 3d band of Fe is more than half filled, the 3d spins of Fe couple antiparallel with the 4f spins of Dy [35, 36]. For heavy rare-earth elements like Dy, spin and orbital moments are oriented parallel. Therefore the 3d moments of Fe and the 4f moments of Dy are aligned antiparallel, resulting in an antiferromagnetic coupling between Fe and Dy [21].

### A. SAMPLE GROWTH

Since the focus of our work is on the magnetic improvement of Fe-based spin probes, we patterned magnetic stripes (see Fig. 1) with dimensions comparable to those of typically used spin injectors (stripe width $W = 1 - 3$ μm, length $L = 400$ μm) on undoped (001) GaAs substrates using electron-beam lithography. After developing the PMMA resist, the substrates were treated with hydrochloric acid (HCl) to remove PMMA residues and the GaAs native oxide. Immediately after that, we mounted the sample in an ultra-high vacuum chamber (base pressure $8 \times 10^{-10}$ mbar) and deposited thin films of Fe or Dy and bilayers of Fe/Dy by magnetron sputtering. The deposition process took place at room temperature and was controlled by time using computer controlled shutters.

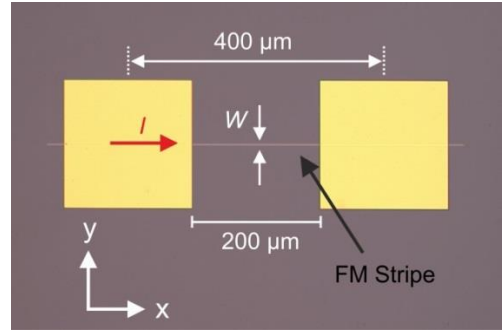

FIG. 1. Micrograph of a microstructured FM stripe (width $W$). A constant dc current $I = 50$ μA is flowing through the FM stripe between two Ohmic contacts. Using the same contacts, the two-terminal magnetoresistance is monitored as a function of the magnetic field, which is applied in the $xy$ plane of the sample.

The Fe single layer was sputtered at low power (5 W) and consists (in order of growth) of 2.9 nm Fe and a cap of 12 nm Au, which prevents the material from oxidation. The Dy single layer consists of either 35 nm Dy or 75 nm Dy and was deposited at high sputtering power (50 W).

The Fe/Dy bilayer was fabricated by deposition of a thin Fe layer on the GaAs substrate, and subsequent deposition of the Dy layer without breaking the vacuum. We manufactured different Fe/Dy bilayers where the thickness of the Dy layer was fixed at 35 nm and the thickness of the Fe layer was increased from 2.5 to 15 nm. After deposition of the FM layers, Ohmic contacts to the stripes were fabricated by electron-beam lithography and thermal evaporation of Ti and Au. The FM stripes are patterned along the [110] direction, however, stripes oriented along the [1$\bar{1}$0] direction yield similar magnetic properties. This suggests that the deposited material is polycrystalline. We also prepared similarly grown Fe, Dy and Fe/Dy full film samples (5 mm x 5 mm) on undoped (001) GaAs substrates for measurements in a super conducting quantum interference device (SQUID) measurement setup.

### B. MEASUREMENT TECHNIQUES

In order to characterize the magnetic properties of the particular FM, we carried out SQUID measurements of full film samples and magnetoresistance measurements of microstructured FM stripes with a similar layer sequence. The hysteresis curves of the full film samples were obtained from measurements using a Quantum Design SQUID setup, which provides fields up to 7 T. The anisotropic magnetoresistance (AMR) effect [37] was used as a tool to determine the orientation of the FM stripe's magnetization axis with respect to the external magnetic field. The basic principle of the AMR effect is that the magnetoresistance of a FM material depends on the relative angle between the magnetization $\mathbf{M}$ and the direction of the electrical current $\mathbf{I}$, which is flowing through the FM material. For a FM it is commonly found that $R_\parallel > R_\perp$, i.e. the resistance $R_\parallel$ is larger when the magnetization axis $\mathbf{M}$ is (anti-) parallel the current direction $\mathbf{I}$, compared to the situation when the magnetization of the FM is transverse to the current direction ($R_\perp$) [38]. The AMR value, which is the relative change of the resistance, is commonly defined as AMR = $(R_\parallel - R_\perp)/R_\perp$.

The magnetoresistance measurements were carried out between 4.2 K and 120 K in a $^4$He cryostat with $\mathbf{B}_{max}$ = 10 T. The samples were mounted in a sample holder which allows rotating the sample in-plane with respect to the magnetic field. We pass a constant dc current $I$ = 50 µA through a single FM stripe by using its Ohmic contacts (see Fig. 1) and monitor simultaneously the two-terminal voltage drop between the Ohmic contacts as a function of the magnetic field. The following measurement routines were used to characterize a particular FM layer sequence.

(1) First, we conduct SQUID measurements with a full film sample, from which we obtain the hysteresis curve of the FM material. Here, the magnetic field is applied parallel to the in-plane direction of the film.

(2) Second, we perform circular magnetic field measurements using microstructured FM stripes with the same layer sequence. A constant magnetic field is rotated in the *xy* plane of the sample (see Fig. 1) by 360°. For FM materials, the characteristic AMR dependence of the resistance, $R(\theta) = R_\perp + (R_\parallel - R_\perp)\cos^2\theta$, is observed. Here, $\theta$ is the angle enclosed by the direction of the current **I** and the magnetization axis **M** of the FM.

(3) By applying a magnetic field in the *xy* plane of the FM stripe, we can determine its coercive field. The magnetic field is swept either (anti-) parallel to stripe (along the *x* direction, what corresponds to $\theta = 0°$ and $\theta = \pm 180°$, respectively) or transverse to the stripe (*y* direction, $\theta = \pm 90°$). By comparing the magnetoresistance data with SQUID data of the corresponding full film samples [see (1)], we can determine the coercive field of the FM stripe.

(4) Finally, we use a measurement routine which is similar to the one employed for SHE experiments described in Sec. I. First, we set the magnetization axis **M** along the *x* axis, i.e. parallel to the stripe's axis. After that, a transverse magnetic field $B_y$ is applied along the *y* axis. By analyzing the obtained magnetoresistance data, we can determine the critical field $B_{x,crit}$ where the FM magnetization has mostly oriented parallel to the applied field, i.e. along the *y* axis. In terms of the previously mentioned SHE experiments, the critical field corresponds the field where the SHE induced signal, detected by the FM spin probes, becomes zero. We also performed measurements where **M** was set transverse to the stripe (*y* axis), and a magnetic field $B_x$ was applied along the stripe's axis (*x* axis).

We first present SQUID and magnetoresistance data of single Dy and Fe layers, and then of Fe/Dy bilayers. By comparing the corresponding results, we are able to distinguish the contributions of the individual FM layer to the combined SQUID and magnetoresistance signal of the Fe/Dy bilayer.

## III. CHARACTERIZATION OF DY SINGLE LAYER

First, we characterize the magnetic properties of thin Dy films. The results presented in the following help to interpret data obtained for Fe/Dy bilayers (see Sec. V.), where similar Dy films were used.

Data on experiments with thin (microstructured) Dy films is very rare in literature. Studies of thin Dy films at low temperatures [39] show that large magnetic fields (5 T) are necessary to orient the film's magnetization. Since the material employed in our studies is presumably polycrystalline, even larger fields might be required to fully saturate the material [24]. Dy stripes with dimensions comparable to our samples have previously been employed to structure magnetic superlattices on 2DEGs [40-42]. These experiments also indicate that magnetic fields of 4-6 T are necessary to orient the magnetization of the Dy superlattice.

### A. MEASUREMENTS AT HELIUM TEMPERATURE

Figure 2(a) shows the SQUID hysteresis loop of a Dy full film sample ($t_{Dy}$ = 75 nm). The curve was recorded at 5 K, where Dy is in the FM phase. The extracted coercive field of the sample is 1150 mT. This is in good agreement with the coercive field reported in Ref. [39] for sputtered, 50 nm thick Dy films, which were deposited at room temperature. The magnetization at 7 T, extracted from SQUID data, is $\mu_0 M$ = 1.95 T, and the remanent magnetization is $\mu_0 M_r$ = 0.76 T. Although the maximum applied field (7 T) is in our experiment larger than in Ref. [39], the magnetization curve does not saturate.

Therefore we applied an even larger magnetic field of 10 T for circular magnetic field sweeps. The magnetoresistance signal of a single Dy stripe ($t_{Dy}$ = 75 nm, $W$ = 2 µm) as a function of the angle $\theta$ (**B**, **I**) is shown in Fig. 2(b). At 0°, **B** is parallel to the Dy stripe (i.e. oriented along the *x* direction, see Fig. 1), and thus parallel to the current **I**, which is flowing through the FM stripe. At ±90° the magnetic field is oriented perpendicular to the stripe (*y* direction). The signal yields a typical AMR signature with a sinusoidal shape of $R(\theta)$. The resistance $R(\theta)$ has a minimum for $\theta = 0°$ ($\mathbf{M}_{Dy} \parallel \mathbf{I}$) and a maximum for $\theta = \pm 90°$ ($\mathbf{M}_{Dy} \perp \mathbf{I}$), i.e. $R_\perp > R_\parallel$. As a consequence, Dy exhibits a negative AMR effect $(R_\parallel - R_\perp)/R_\perp < 0$ with an AMR value of roughly -0.2 %. Previously, a negative sign of the AMR effect has only been reported for certain alloys or compounds, e.g. for (Ga,Mn)As [43, 44], Ni- and Mn-based alloys [45, 46], Fe$_4$N [47] or half-metallic ferromagnets [48], but not for rare-earth elements.

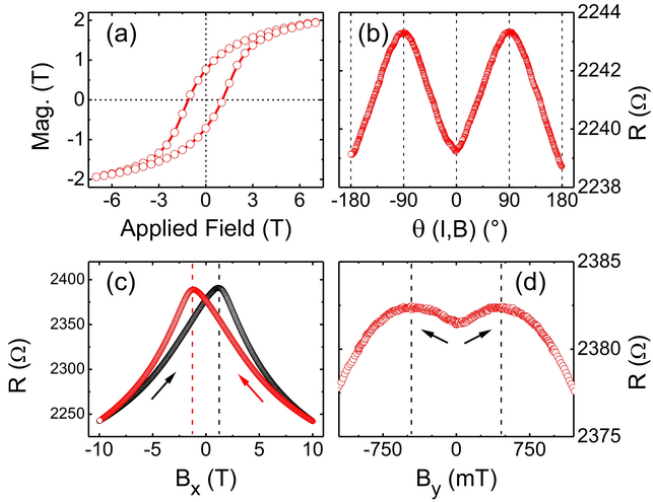

FIG. 2. (a) SQUID hysteresis loop of a Dy full film sample ($t_{Dy}$ = 75 nm), obtained at 5 K. (b) Magnetoresistance of a single Dy stripe (75 nm Dy, $W$ = 2 µm) at 4.2 K as a function of the angle between magnetic field **B** and current **I**. The constant magnetic field (10 T) was rotated in the $xy$ plane of the sample. **B** is parallel to the stripe and to the current **I** at 0° ($x$ direction) and perpendicular at ±90° ($y$ direction). (c) Data of coercive field measurements at 4.2 K, with $B_x$ oriented parallel to the stripe. Dashed lines mark the coercive field. (d) Critical field measurement of the same stripe at 4.2 K. The magnetization of the stripe was set along the $x$ direction (along the stripe's axis) before a perpendicular field $B_y$ was applied. The critical field is indicated by dashed lines.

A quantitatively similar behavior was obtained for Dy stripes with different widths ($W$ = 3 µm and 1 µm). We also deposited Dy stripes with a different layer thickness (35 nm) on GaAs and on a Si/SiO$_2$ substrate, and we additionally capped some of the samples with a thin layer of Au (12 nm) to prevent the surface from oxidization. Magnetoresistance data of these samples also reveal a negative sign of the AMR effect, ruling out that the effect derives either from the substrate or from surface oxidation. For smaller magnetic fields (i.e. 6 T) the $R(\theta)$ shape does not show a regular AMR $\sin^2(\theta)$ signature. This suggests, that the magnetization axis does not instantly reorient along **B** for small magnetic fields, emphasizing the hard FM character of Dy.

From the results described above we can assume that $\mathbf{M}_{Dy}$ is oriented along the **B**-field vector for **B** = ±10 T. We now apply the magnetic field either parallel to the stripe ($B_x$) or transverse to the stripe ($B_y$), and sweep it from 10 T to -10 T and back. This allows us to investigate the coercive field of the Dy stripe along the $x$- and $y$ direction. The magneto-resistance data for the $B_x$ sweep is depicted in Fig. 2 (c). When the magnetic field is ramped down from 10 T to -10 T (red curve), the slope of the magnetoresistance reverses its sign at $B_x$ = -1167 mT, when the signal is maximum. The curve of the upsweep (-10 T to 10 T, black curve) is mirror-symmetric, and the signal is maximum at $B_x$ = +1167 mT.

The decrease of the magnetoresistance with increased $|B_x|$ can be explained by the negative magnetoresistance effect (NMR), which is commonly observed for FM materials [49]. However, by comparing the magnetoresistance data with corresponding SQUID data of the full film sample [see Fig. 2 (a)] we find that the maximum of the signal / reversal of the slope at $|B_x|$ = 1167 mT coincidences with the coercive field, which we extracted from SQUID data (1150 mT). We assume that the magnetization reversal leads to an increased spin disorder and thus to an increase of the magnetoresistance. The spin disorder is maximal at the coercive field and decreases, when the Dy stripe reverses its magnetization direction. Thus, the maximum of the signal / reversal of the slope corresponds to the coercive field $|B_{x,coerc}|$ = 1167 mT of the Dy stripe, which is also marked with dashed lines in Fig. 2(c).

The curve obtained from the magnetic field sweep along the $y$ direction exhibits a similar trace (not shown here), the reversal of the slope of the $B_y$ curve, however, occurs at a lower magnetic field ($|B_{y,coerc}|$ = 953 mT). This finding can be understood in terms of shape anisotropy [50], by which the long axis of the stripe ($x$ axis) is magnetically preferred over the shorter axis ($y$ axis).

Finally, we apply the SHE-like measurement routine, which was described in Sec. II., to the Dy stripe, We use a field of $B_x$ = 10 T to set the magnetization $\mathbf{M}_{Dy}$ along the stripe's axis, i.e. along the $x$ direction. When the field is ramped back to zero, the magnetization of the stripe will stay mainly aligned in this direction, since the employed Dy film exhibits a large coercive field and a finite remanence. After that we sweep the magnetic field $B_y$ transverse to $\mathbf{M}_{Dy}$, i.e. along the $y$ direction, from zero field to +10 T or -10 T. By analyzing the magnetoresistance data, we can determine the critical field $B_{x,crit}$, where the FM magnetization orients preferentially parallel to the applied field. The corresponding data is shown in Fig. 2(d). The magnetoresistance exhibits a local minimum at zero field and increases monotonously until $B_y \approx \pm 500$ mT. For $|B_y|$ > 500 mT, the slope of the curve reverses its sign, so that the magnetoresistance decreases with increasing $|B_y|$. From these data we can extract the critical field. At $B_y$ = 0 mT, the magnetization of the stripe is oriented along the stripe's axis ($x$ direction), so that $\mathbf{M}_{Dy} \parallel \mathbf{I}$. Since Dy exhibits a negative AMR effect, the AMR signal is minimum at zero field. However, when $B_y$ is increased, $\mathbf{M}_{Dy}$ will start to reorient along the $y$ axis with the applied magnetic field. When $\mathbf{M}_{Dy}$ is largely oriented parallel to $B_y$, i.e. perpendicular to the stripe, the AMR induced signal yields a maximum ($\mathbf{M}_{Dy} \perp \mathbf{I}$).

The decrease of the magnetoresistance for $|B_y| > \pm 500$ mT is again due to the NMR effect. Overall, the recorded Dy magnetoresistance curve is a superposition of the AMR and NMR induced signals. The position of the maximum of the magnetoresistance curve is now defined as the critical field $B_{crit}$ and is determined from the signal's derivative. Here, we extract a critical field of $|B_{x,crit}| = 445$ mT, marked with dashed lined in Fig. 2(d).

We also performed measurements where the initial orientation of $\mathbf{M}_{Dy}$ was set transverse to the stripe ($y$ direction) and a magnetic field $B_x$ was swept along the $\pm x$ direction, i.e. parallel to the stripe. Here, we observe a transition of the signal (not shown) from a maximum to a relative minimum of the signal, which is consistent within the AMR effect. The initial orientation of $\mathbf{M}_{Dy}$ along the $y$ direction leads to a maximum of the AMR induced signal ($\mathbf{M}_{Dy} \perp \mathbf{I}$), whereas the final state, when $\mathbf{M}_{Dy}$ is oriented parallel to the stripe, corresponds to a minimum of the AMR signal ($\mathbf{M}_{Dy} \parallel \mathbf{I}$ or $\mathbf{M}_{Dy} \uparrow\downarrow \mathbf{I}$). The extracted critical field $|B_{y,crit}| = 247$ mT is lower in this configuration than for the previous setup. Again, this can be explained by shape anisotropy. Both measurements show that the reorientation of $\mathbf{M}_{Dy}$ under the influence of a transverse magnetic field can consistently be explained and traced by means of the AMR effect.

## B. TEMPERATURE DEPENDENT MEASUREMENTS

We also studied the temperature dependence of the coercive field and of the critical field between 4.2 K and 120 K by SQUID and magnetoresistance measurements. With increasing temperature, the coercive field (Fig. 3, squares and stars) and the critical field (Fig. 3, circles) decrease. This is ascribed to the decrease of the magnetocrystalline anisotropy of the Dy layer with increasing temperature [51]. Again, the coercive field extracted from SQUID measurements (Fig. 3, stars) and from magnetoresistance measurements (Fig. 3, squares) are in good agreement. At $T = 15$ K, the coercive field has dropped to $|B_{x,coerc}| = 900$ mT, whereas $|B_{x,coerc}| = 270$ mT at 90 K.

Surprisingly we observe a non-zero coercive field (70 mT) and critical field (36 mT) at $T = 120$ K where Dy is supposed to be in the antiferromagnetic phase ($T_C = 90$ K) [29-32], i.e. the coercive and the critical field should be zero. Correspondingly, the circular magnetic field measurements show a sinusoidal AMR signature up to the highest temperature of 120 K. Both findings strongly suggest, that Dy has still a FM component at $T = 120$ K.

To investigate the FM-AFM phase transition in more detail, we performed two different kinds of SQUID measurements with a Dy full film sample ($t_{Dy} = 75$ nm Dy).

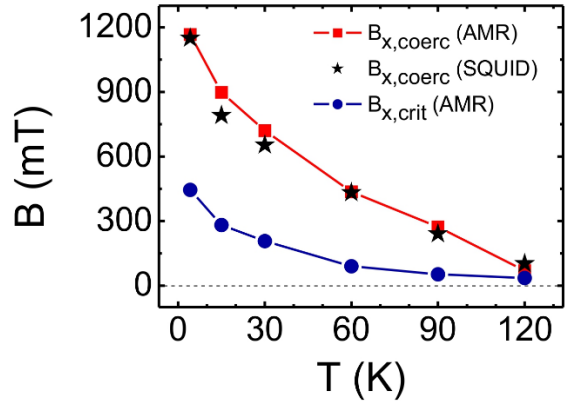

FIG. 3. Temperature dependence of the coercive field $B_{x,coerc}$ (square) and of the critical field $B_{x,crit}$ (circles) of a single Dy stripe (75 nm Dy, $W = 2$ μm). The coercive field extracted from SQUID data (stars) is obtained from a similarly grown full film sample. Lines are guides for the eye only.

First, we recorded the magnetization curve during field cooling (FC) from 300 K to 5 K with an applied field of 7 T. The recorded magnetization of the FC curve, shown in Fig. 4 (a), becomes non-zero below 160 K. This indicates that the Dy film has a FM component up to 160 K, i.e. way above the Curie temperature of crystalline Dy.

A similar result was also obtained from the second SQUID experiment. Here, the sample was zero field cooled (ZFC) and the magnetization was oriented at 5 K in the plane of the sample with a field of 7 T. Afterwards, the sample was warmed up to 300 K in zero field (ZF) while the sample's remanent magnetization was recorded as a function of temperature. The warming curve [see Fig. 4(a)] clearly shows that the FM-AFM phase transition is shifted to 160 K, since the recorded magnetization becomes zero only for $T > 160$ K. Both experiments confirm that sputtered Dy has a FM component even above its nominal Curie temperature and that the AFM-FM magnetic phase transition occurs at a higher temperature ($\approx 160$ K) than reported for single crystalline Dy [29-32].

In addition, we recorded the resistance of a single Dy stripe ($t_{Dy} = 75$ nm, $W = 2$ μm) during ZFC from room temperature to 4.2 K ($I = 50$ μA). As expected for metals, a decrease of the resistance is observed during cooling [see Fig. 4(b)]. At $T \approx 169$ K, an anomaly of the resistance, marked by dashed line in Fig. 4(b), is visible and indicates a magnetic phase transition of the Dy stripe about the same temperature at which the magnetization vanishes in SQUID experiments.

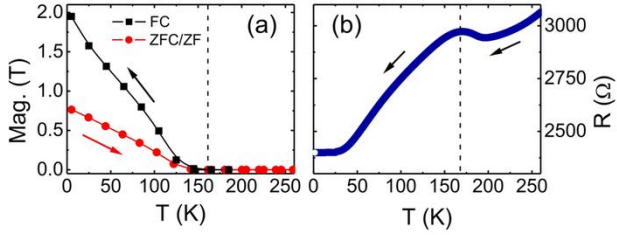

FIG. 4. (a) FC (squares): SQUID magnetization curve of a full film Dy sample (75 nm Dy), recorded during field cooling (applied field = 7 T) from 300 K to 5 K. ZFC/ZF (circles): Zero field warming curve for the same sample. Prior to heating, the sample was zero field cooled to 5 K and the magnetization was oriented in the plane of the sample with a field of 7 T. Both FC and ZFC/ZF curves exhibit a magnetic phase transition around 160 K (dashed line). (b) Electrical resistance of a single Dy stripe (75 nm Dy, $W$ = 2 μm) as a function of temperature. The stripe was zero field cooled from room temperature to 1.4 K. The anomaly in the resistance (≈169 K, marked with a dashed line) indicates a magnetic phase transition.

An absence of the FM/AFM phase transition at $T$ = 90 K has also been reported by Beach *et al.* [52, 53] for Dy lattices grown on Lutetium layers. They found that the compressive epitaxial strain between the Lutetium layers and the Dy lattice leads to an enhancement of the Curie temperature of Dy up to $T$ = 175 K. A similar finding was also made by Scheunert *et al.* [39] for sputtered Dy films, which were deposited at room temperature. They conclude that the deposition at room temperature induces strain in the hcp grain lattice of Dy, leading to a suppression of the FM-AFM phase transition at 90 K and a shift of the Curie temperature up to 172 K. Since the growth conditions of our samples (sputtered at room temperature) are comparable to those of Scheunert *et al.* [39], it is most likely that the shift or suppression of the magnetic phase transition is induced by strain in the Dy lattice.

Overall, the hard FM characteristic of the employed thin Dy layers is emphasized by the large values of the coercive and the critical field in $x$ as well as in $y$ direction. This shows that Dy can be expected to establish in-plane magnetization states which are robust against external magnetic field.

## IV. CHARACTERIZATION OF FE SINGLE LAYER

We now characterize the magnetic properties of thin Fe films. The data helps to evaluate the magnetic interplay between the Fe and the Dy layer in Fe/Dy bilayers, discussed in Sec. V.

Figure 5(a) shows the SQUID hysteresis loop (downsweep) of a Fe full film sample (2.9 nm Fe / 12 nm Au), which was recorded at 5 K. The coercive field of the Fe sample is small (2.6 mT), characteristic for the small magnetocrystalline anisotropy of the Fe layer. On the other hand, the Fe layer exhibits a large magnetization at 7 T of $\mu_0 M$ = 2.0 T and a finite remanent magnetization of $\mu_0 M_r$ = 0.99 T. Both features, a small coercive field and a large saturation magnetization, are typically found for soft FMs [18].

The microstructured Fe stripes (2.9 nm Fe / 12 nm Au, $W$ = 3 μm) were first characterized by circular magnetic field measurements at 4.2 K, as shown in Fig. 5(b). Since the Fe layer's magnetization saturates at 40 - 50 mT [see. Fig 5(a)], a field of $\mathbf{B}$ = 1 T was applied. The circular magnetic field measurements show that the AMR induced signal is largest at $\theta$ = 0° when current and magnetization are parallel ($\mathbf{M}_{Fe} \parallel \mathbf{I}$), and is smallest at $\theta$ = ±90° when $\mathbf{M}_{Fe} \perp \mathbf{I}$.

Therefore, Fe exhibits (in contrast to Dy) a positive AMR effect with a characteristic $\cos^2(\theta)$ dependence and an AMR value of roughly 0.1 %. This finding allows us to distinguish between the contributions of Fe (positive AMR effect) and Dy (negative AMR effect) to the magnetoresistance signal of the Fe/Dy bilayer (see Sec. V.)

The coercive field of the Fe stripe is determined by sweeping $B_x$ in the range of ±1 T, as shown in Fig. 5(c). When the magnetic field is ramped down from 1 T (red curve), the magnetoresistance increases monotonously until $B_x \approx$ +10 mT is reached. Between +10 mT and -30 mT, a change of the magnetoresistance from maximum to a relative minimum ($B_x$ = - 9 mT, marked with red dashed line) and back, can be noticed. For $B_x$ > -30 mT, the magnetoresistance decreases monotonously with $B_x$. The magnetoresistance signal of the upsweep (-1 T to 1 T, black curve) is mirror-symmetric, and the signal becomes minimum at $B_x$ = +9 mT (marked with black dashed line).

In the following, we analyze the trace of the downsweep curve (+1 T to -1 T). The decreasing magnetoresistance with $|B_x|$ for fields larger than ±30 mT is due to the NMR effect. However, the change in the magnetoresistance between +10 mT and -30 mT cannot be described by the NMR effect, since the NMR effect scales monotonously with the magnetic field. Here, the change in the signal is related to the reorientation of the stripe's magnetization along the reversed magnetic field direction and can be explained by means of the AMR effect.

The reorientation of the stripe's magnetization from +$x$ to –$x$ direction can be mediated by 180° domain wall motion and rotation of the magnetization axis. If reorientation of the magnetization occurs through 180° domain wall motion only, the AMR signal remains unchanged, since it scales as $\cos^2 \theta$, i.e. the parallel and antiparallel orientation of the magnetization axis are equivalent. If the reorientation occurs through rotation of the magnetization axis in the plane of the sample, the AMR signal changes with the angle $\theta$ between current $\mathbf{I}$ and the magnetization axis.

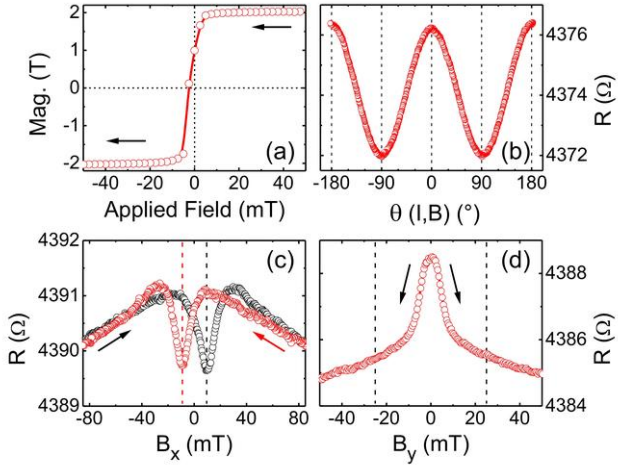

FIG. 5. (a) SQUID hysteresis loop (downsweep) of a 2.9 nm Fe full film sample, obtained at 5 K. (b) Magnetoresistance of a single Fe stripe (75 nm , $W$ = 3 μm) at 4.2 K as a function of the angle between the magnetic field **B** and the current **I**. The constant magnetic field (1 T) was rotated in the $xy$ plane of the sample. (c) Coercive field measurements at 4.2 K with $B_x$ oriented parallel to the stripe's axis. Dashed lines mark the coercive fields. (d) Critical field measurements of the same stripe at 4.2 K. $B_{x,crit}$ is marked with dashed lines.

This explains consistently the change of the magneto-resistance between +10 mT and -30 mT and determines the coercive field of the stripe.

For large and positive $B_x$, the stripe's magnetization $\mathbf{M}_{Fe}$ is oriented parallel to the magnetic field and thus to the stripe. When the magnetic field reverses sign, $\mathbf{M}_{Fe}$ starts to realign along the reversed magnetic field. The magnetization rotates in the plane of the sample from its initial orientation (along the $+x$ direction, $\mathbf{M}_{Fe} \parallel \mathbf{I}$) towards the $y$ axis, and then towards the reversed magnetic field direction, so that $\mathbf{M}_{Fe}$ is finally aligned antiparallel to **I** in $-x$ direction.

In terms of the AMR signal, the rotation of $\mathbf{M}_{Fe}$ corresponds to a change from a maximum ($\mathbf{M}_{Fe} \parallel \mathbf{I}$), to a minimum of the signal ($\mathbf{M}_{Fe} \perp \mathbf{I}$), and finally back to a maximum ($\mathbf{M}_{Fe}$ is aligned along the $-x$ direction, $\mathbf{M}_{Fe} \Updownarrow \mathbf{I}$). With that, we can determine the coercive field of the Fe stripe, given by the AMR minimum. The coercive field extracted from the downsweep curve is $B_{x,coerc}$ = -9 mT [marked with red dashed line in Fig. 5 (b)], and the coercive field for the upsweep is equally +9 mT (black dashed line).

Coercive fields of the same order of magnitude have also been observed in spin injection experiments with Fe spin probes of comparable thickness [1, 2, 7]. The coercive field of the Fe stripe, which we obtained from magnetoresistance measurements, is larger than the coercive field obtained from SQUID data [Fig. 5(a), 2.6 mT]. This is due to shape anisotropy (stripe vs. rectangular sample), which is more dominant for soft FMs due to the small magnetocrystalline anisotropy.

The magnetoresistance data of the Fe stripe also provide information on the magnetization reversal process. By comparing the circular magnetic field measurement [Fig. 5(b)] and the coercive field measurements, we find that the change in the AMR signal in the former case is larger ($\Delta R_{circular}$ = 4.2 Ω) than in the latter case ($\Delta R_{coerc}$ = 1.5 Ω). For the circular magnetic field measurement, the magnetization axis follows instantly the applied field, so that it rotates coherently without domain wall motion. Therefore, the change in the AMR signal is at maximum. On the other hand, the change of the AMR signal observed for the coercive field measurement is significantly smaller. This suggests that the magnetization reversal is not solely mediated by rotation of the magnetization axis, but also through 180° domain wall motion, which does not contribute to a change of the AMR signal. The reorientation of the Fe stripe's magnetization is therefore mediated by incoherent rotation of the magnetization axis and by 180° domain wall motion.

Finally, we apply the SHE-like measurement routine to the Fe stripe. This is shown in Fig. 5(d). Here, a field of $B_x$ = 1 T is sufficient to set the magnetization $\mathbf{M}_{Fe}$ along the stripe's axis (i.e. along the $x$ direction). When the field is ramped back to zero, the magnetization of the stripe will mainly stay aligned along its axis. The magnetic field $B_y$ is then swept transverse to $\mathbf{M}_{Fe}$, i.e. along the $y$ direction, from zero field to +1 T or -1 T. The trace of the signal is again related to the rotation of the Fe stripe's magnetization axis with the applied magnetic field. The recorded curve, shown in Fig. 5(d), exhibits a distinct maximum at $B_y$ = 0 T and drops rapidly between $B_y$ = ±25 mT. For $|B_y|$ > 25 mT the curve becomes linear due to the NMR effect.

The trace of the recorded signal can be explained by means of the AMR effect. At zero field, $\mathbf{M}_{Fe}$ is aligned parallel to the stripe, so that the AMR signal is maximum ($\mathbf{M}_{Fe} \parallel \mathbf{I}$). When $B_y$ is increased, the magnetization orients parallel to the applied magnetic field and thus transverse to the stripe. This state corresponds to a minimum of the AMR signal ($\mathbf{M}_{Fe} \perp \mathbf{I}$). The critical field is now derived from the analysis of the curve and its derivative. When $\mathbf{M}_{Fe}$ has oriented mostly parallel to the applied field (i.e. $\mathbf{M}_{Fe} \perp \mathbf{I}$), the AMR signal is minimum and the slope is zero. If only the AMR effect would contribute to the total magnetoresistance, the critical field is equal to the position, where the derivative of the signal becomes zero. Nevertheless, we also have to take into account the NMR signal, which is dominant for $|B_y|$ > 25 mT. Thus, the critical field is equal to the position, where the derivative approaches a finite, constant value, i.e. where the AMR slope is zero and only the constant NMR slope contributes to the derivative. By analyzing the derivative of the total signal and by fitting the NMR induced background (not shown), we can extract the value of the critical field, $|B_{x,crit}| = 25$ mT, which is marked with dashed lines in Fig. 5(d).

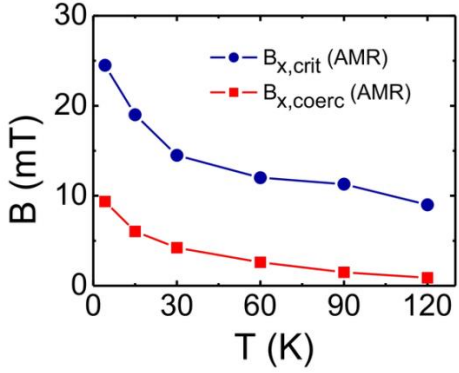

FIG. 6. Temperature dependence of the coercive field $B_{x,coerc}$ (squares) and of the critical field $B_{x,crit}$ (circles) of a single Fe stripe (2.9 nm Fe, $W = 3$ μm). Both curves were obtained from magnetoresistance measurements. Lines are guides for the eye.

The temperature dependence of the coercive field and the critical field was studied in the same temperature range as for the Dy single layer (4.2 K - 120 K) and is shown in Fig. 6. A decrease of $B_{coerc}(T)$ (squares) and $B_{crit}(T)$ (circles) with $T$ can be noticed, whereby the decrease is largest between 4.2 K and 30 K. At $T = 15$ K, the coercive field and the critical field have dropped to $|B_{x,coerc}| = 6$ mT and $|B_{x,crit}| = 19$ mT. Only small values for both fields, $|B_{x,coerc}| = 0.9$ mT and $|B_{x,crit}| = 9$ mT, were recorded at $T = 120$ K. We attribute these findings not only to a decreasing magnetocrystalline anisotropy of the Fe film with increasing temperature, but also to domain wall motion, which is thermally activated.

## V. CHARACTERIZATION OF FE/DY BILAYER

The basic motivation of this work is combining Fe with Dy in Fe/Dy bilayers to pin the soft FM Fe layer by the hard magnetic Dy film. We first present data at helium temperature for a particular bilayer (2.9 nm Fe / 35 nm Dy, $T = 4.2$ K) and figure out which features arise from the interaction between both materials. Then, we study for the same bilayer the temperature dependence of the magnetic interplay between 4.2 K and 120 K. Finally, we characterize different Fe/Dy bilayers ($T = 4.2$ K), where the thickness of the Fe layer $t_{Fe}$ was increased from 2.5 nm to 15 nm while the thickness of the Dy layer was fixed at 35 nm. By varying $t_{Fe}$ for a given thickness of the hard FM Dy layer, we can systematically study the magnetic interplay between both materials. All SQUID measurements are accompanied by magnetoresistance data acquired from similarly grown full film samples.

### A. MEASUREMENTS AT HELIUM TEMPERATURE

The SQUID hysteresis loop of a Fe/Dy full film sample (2.9 nm Fe / 35 nm Dy) was recorded at 5 K and is depicted in Figure 7(a). Although the sample consists of two different magnetic layers, the overall shape of the curve resembles that of a uniform, single phase magnet, thus clearly indicating exchange coupling between both layers [18]. The magnetization curve shows no saturation at 7 T, similar to the one observed for Dy [see Fig. 2(a)].

The degree of coupling between both layers depends on the thickness $t_s$ of the soft FM layer [18]. If the thickness of the soft magnetic material is below a critical value, both FM layers are rigidly coupled and reverse their magnetization at the same field. On the other hand, if $t_s$ is larger than the critical thickness, the soft FM layer reverses its magnetization at fields smaller than the hard FM layer.

The coercive field of the Fe/Dy bilayer (225 mT), which we extract from the SQUID curve in Fig. 7(a), is lower than that of the Dy layer [1150 mT, see Fig. 2(a)]. This suggests that the latter case applies for our sample and that both layers are not rigidly coupled. On the other hand, it can also be noted that the coercive field is considerably larger than that of the Fe sample [2.6 mT, see Fig. 5(a)].

For a soft FM sandwiched between two hard FMs, it was found that the critical thickness is roughly twice the width of the domain wall $\delta_h$ in the hard FM layer [20, 54]. For the layer sequence employed in our sample, the critical thickness is equal to $\delta_h$. Calculations by Egami and Graham [55] yield a Dy domain wall thickness of about 7 atomic layers (at zero temperature). This value corresponds to $\delta_{Dy} \approx 2$ nm and is in line with the experimental observations, which suggest that the Fe layer thickness ($t_{Fe} = 2.9$ nm) is larger than the critical thickness.

In the following, we characterize a microstructured Fe/Dy bilayer, which has the same layer sequence as the sample used for SQUID measurements (2.9 nm Fe / 35 nm Dy, $W = 2$ μm). The magnetoresistance curves yields distinct features, which have been recorded neither for the Dy single layer, nor for the Fe single layer. The analysis of the obtained magnetoresistance data helps to interpret the SQUID magnetization curves and to understand the interplay between the Fe and the Dy layer.

First, we elucidate how the Fe/Dy bilayer compares electrically to the Fe and the Dy single layers. For that, we treat the Fe/Dy bilayer as a conductor, where the Fe and the Dy layer form a parallel circuit. The ratio of the currents $I_{Dy}$ and $I_{Fe}$, which are flowing through the respective layer of the bilayer, is then given by

$$\frac{I_{Fe}}{I_{Dy}} = \frac{R_{Dy}}{R_{Fe}}. \qquad (1)$$

The resistances $R_{Fe}$ and $R_{Dy}$ were determined by measuring the four-terminal resistance of the respective Fe and Dy single layer samples. For the bilayer discussed in the following (2.9 nm Fe / 35 nm Dy) we found, that the resistance of the Dy layer is approximately 2.5 times larger than that of the Fe layer.

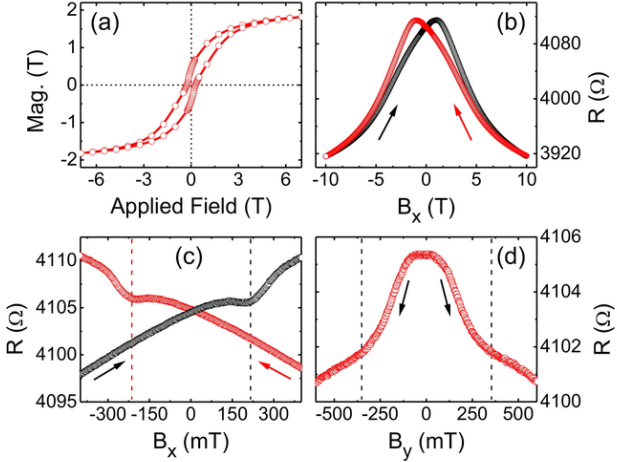

FIG. 7. (a) SQUID hysteresis loop of a full film Fe/Dy bilayer (2.9 nm Fe / 35 nm Dy), obtained at 5 K. (b) Magnetoresistance data of coercive field measurements at 4.2 K of a single Fe/Dy stripe ($W = 2$ μm) with the same layer sequence. (c) Zoom of the data given in (b). Dashed lines mark the coercive field of the exchange coupled Fe layer. (d) Critical field measurement of the same stripe at 4.2 K. The critical field $B_{x,crit}$ of the exchange coupled Fe layer is marked with dashed lines.

In consequence, the current $I_{Fe}$ flowing through the Fe layer is 2.5 times larger than the current $I_{Dy}$ flowing through the Dy layer. Thus, ≈71 % of the total current in the Fe/Dy bilayer is flowing through the Fe layer and ≈29 % through the Dy layer.

The coercive field of the bilayer was determined by sweeping $B_x$ in the range of ±10 T. The obtained curves, shown in Fig. 7(b), resemble qualitatively that of the Dy single layer [see Fig. 2(b)]. When the magnetic field is ramped down from 10 T to -10 T (red curve), the magnetoresistance first increases monotonously due to the NMR effect and the slope of the signal is continuous around zero field. The slope reverses its sign at $B_x = -1043$ mT, when the signal is maximum. Then the magnetoresistance decreases monotonously with $B_x$ for increasing negative magnetic field. The curve of the upsweep (-10 T to 10 T, black curve) is mirror-symmetric, and the signal is maximum at $B_x = +1043$ mT. This suggests that the magnetic properties of the bilayer are mainly determined by the hard FM Dy layer. Therefore we apply the same analysis as for the single layer Dy stripe (Sec. III.). With that, we determine the coercive field of the Dy layer, $|B_{x,coerc} (Dy)| = 1043$ mT, which is comparable to the value obtained for the single Dy layer (1167 mT).

However, zooming into the curve, shown in Fig. 7(c), reveals, e.g. for the downsweep (red curve) a relative minimum of the signal at $B_x = -217$ mT (equally at $B_x = +217$ mT for the upsweep, black curve). Such feature has not been observed in the $B_x$ data for the Dy single layer. On the other hand, a similar trace of signal has been recorded for the coercive field measurements of the microstructured Fe single layer [see Fig. 5(c)].

If the NMR induced background is fitted and subtracted (not shown), a transition of the magnetoresistance from a maximum to a relative minimum is visible. This suggests, that the feature is related to the Fe layer. Analogous to the analysis of the single Fe layer (see Sec. IV.), the trace of the signal can be explained by the AMR effect.

For large and positive $B_x$, the magnetization of the Fe layer is aligned parallel to the magnetic field and thus parallel to the stripe's axis. When $B_x$ reverses its sign, $\mathbf{M_{Fe}}$ rotates from its initial orientation (along the stripe's axis, $+x$ direction) towards the $y$ axis (i.e. perpendicular to the stripe's axis) and finally towards the reversed magnetic field direction ($-x$ direction). In terms of the Fe AMR effect, this corresponds to a change of the AMR induced signal from a maximum ($\mathbf{M_{Fe}} \parallel \mathbf{I}$) to a minimum ($\mathbf{M_{Fe}} \perp \mathbf{I}$), and finally back to a maximum ($\mathbf{M_{Fe}} \Updownarrow \mathbf{I}$). We can exclude that the observed change of the signal stems from the rotation of the Dy layer's magnetization instead, since this would result in a reversed signal shape due to the negative sign of the Dy AMR effect. With that, we extract the enhanced coercive field of the Fe layer, $|B_{x,coerc} (Fe)| = 217$ mT, which is marked with dashed lines in Fig. 7(c). For magnetic field sweeps along the $y$ direction, we could consistently observe a change in the signal (not shown) from a minimum to a maximum and back.

Analyzing the magnetoresistance curve allows thus distinguishing between the magnetization reversal of the Fe and of the Dy layer. Furthermore, the magnetoresistance data also verifies, that the magnetization reversal and the coercive field observed in the SQUID curve [Fig. 7(a)] can be attributed to the Fe layer. Moreover, the coercive fields extracted from SQUID [225 mT in Fig. 7(a)] and magnetoresistance data ($|B_{x,coerc}| = 217$ mT) are in good agreement.

All in all, we observe that the Fe layer is coupled by the Dy layer and that the Fe layer's coercive field is significantly enhanced by a factor of more than 24, compared to the Fe single layer ($|B_{x,coerc}| = 9$ mT). However, the Fe layer reverses its magnetization at fields smaller than the hard FM Dy layer, whereas the Dy layer's coercive field remains nearly unchanged. This finding supports our assumption that the Fe thickness is above the critical thickness, so that the coupling between both layers is not perfectly rigid.

We also performed measurements with the Fe/Dy bilayer, from which we can determine the critical field. In this measurement routine we use a field of $B_x = 10$ T to align the bilayer's magnetization along the stripe's axis ($x$ direction). The magnetic field is then ramped back to zero. Afterwards, the magnetic field $B_y$ is swept transverse to the stripe's axis from zero field to either $B_y = +10$ T or to $B_y = -10$ T.

Corresponding data is shown in Fig. 7(d), exhibiting a broad maximum at $B_y = 0$ T, followed by a rapid decrease of the magnetoresistance down to $|B_y| \approx 350$ mT. For $|B_y| > 350$ mT the trace of the curve becomes roughly linear. Comparison of the Fe/Dy bilayer signal with that of the single Fe layer [see Fig. 5(d)] suggests that the observed trace is again connected with the rotation of the Fe layer's magnetization. Similar to the Fe layer stripe, the trace of the recorded signal can be explained by means of the AMR effect. At zero field, $\mathbf{M}_{Fe}$ is aligned parallel to the stripe, so that the AMR induced signal shows a maximum ($\mathbf{M}_{Fe} \parallel \mathbf{I}$). For large $B_y$, the Fe layer's magnetization has mostly oriented parallel to the applied magnetic field and thus transverse to the stripe. This state corresponds to a minimum of the AMR signal, since $\mathbf{M}_{Fe} \perp \mathbf{I}$. Again, we can rule out that the signal stems from the Dy layer's magnetization instead, because of the negative sign of the Dy layer's AMR effect.

The critical field of the Fe layer is derived in the same manner as before. When $\mathbf{M}_{Fe}$ has oriented mostly parallel to the applied field (i.e. $\mathbf{M}_{Fe} \perp \mathbf{I}$), the AMR induced signal is minimum, and the signal's derivative becomes zero. Nevertheless, we also have to take into account the NMR induced signal, which is dominant for $|B_y| > 350$ mT, and which contributes as a constant offset to the derivative of the total signal. With that, we can extract the critical field of the Fe layer, which is marked with dashed lines in Fig. 7(d). The analysis of the total signal's derivative yields $|B_{x,crit}(\text{Fe})| = 352$ mT.

In the reversed configuration, the magnetization is set along the $y$ direction and a transverse field is applied along the stripe's axis ($x$ direction). We observe a transition of the signal from a relative minimum at zero field to a maximum of the signal (not shown here), whereby the trace of the signal is again in accordance with the sign of the Fe AMR effect. Compared to the Fe single layer ($|B_{x,crit}(\text{Fe})| = 25$ mT), the critical field of the Fe layer is significantly enlarged by a factor of more than 14.

Altogether, SQUID and magnetoresistance measurements have clearly shown that both the coercive and the critical field of the Fe layer are increased by more than one order of magnitude, when the Fe layer is brought in contact with the hard FM Dy layer. This strongly suggests that the magnetic properties of the Fe layer are enhanced through soft/hard FM exchange coupling at the interface between both layers.

## B. TEMPERATURE DEPENDENCE

We also performed temperature dependent SQUID and magnetoresistance measurements with the same bilayer (2.9 nm Fe / 35 nm Dy). The temperature was varied between 4.2 K and 120 K. This allows us to further characterize the exchange coupling between both layers, since the magnetic properties of the Fe and the Dy layer change with temperature [see Figs. 3 and 6].

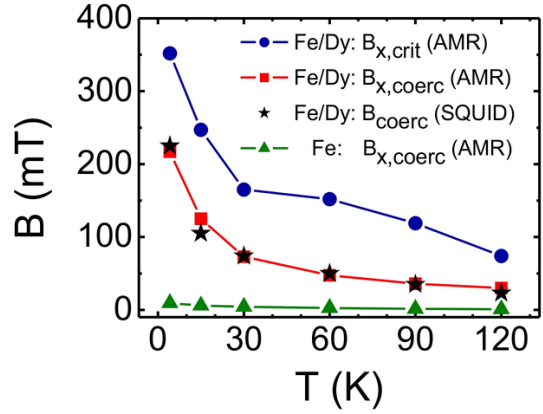

FIG. 8. Temperature dependent coercive and critical fields of the exchange coupled Fe layer in a Fe/Dy bilayer (2.9 nm Fe / 35 nm Dy). Data for $B_{x,coerc}$ (squares) and $B_{x,crit}$ (circles) was obtained from a microstructured FM stripe ($W = 2$ μm). SQUID data of the coercive field (stars) was obtained from a similarly grown full film sample. Lines are guides for the eye.

The coercive field of the microstructured Fe/Dy bilayer was again determined by sweeping the magnetic field between ±10 T. The shape of the obtained curves (not shown) resembles for all temperatures those of single layer Dy. Moreover, at all temperatures the extracted coercive fields of the Dy layer are similar to those of the single layer Dy. This implies that the Dy layer is also at $T = 120$ K in the FM phase (for the same reason as the Dy single layer) and exhibits no FM/AFM transition at 90 K. The interplay between Fe and Dy is therefore determined by FM/FM coupling also at $T = 120$ K. At all temperatures, we record features, which have similarly been observed at helium temperature. This allows us to clearly identify the rotation of the Fe layer's magnetization and thus its coercive field. We also performed temperature dependent SQUID measurements with a similarly grown full film sample. The coercive fields, which we extracted from the magnetization curve, are in good agreement with the coercive fields obtained from the magnetoresistance data.

The temperature dependence of the critical field was determined by applying the same measurement procedure as for 4.2 K. At all temperatures, the $B_y$ curves (pre-magnetization along the $x$ direction, i.e. along the stripe's axis) show a transition of the signal from a maximum at zero field to a relative minimum, as it was observed at 4.2 K. For the $B_x$ measurements (premagnetization is set along the $y$ axis) we consistently observe a change of the signal from a maximum to a relative minimum (not shown). At all temperatures, the critical field of the Fe layer can clearly be identified.

Figure 8 shows the temperature dependence of the coupled Fe layer's coercive field and critical field. The coercive field, which was obtained from SQUID (stars) and from magnetoresistance measurements (squares), decreases rapidly between 4.2 K and 30 K, whereas only a modest decrease is visible between 30 K and 120 K. However, at all temperatures, the Fe layer's coercive field is enhanced by a factor of 14-20, compared to the coercive field of the Fe single layer (triangles). The critical field (circles) of the coupled Fe layer also decreases rapidly between 4.2 K and 30 K, and an almost linear decay can be noticed for higher temperatures. However, at $T$ = 120 K, the Fe layer's critical field is still enhanced by a factor of 8, compared to the Fe single layer (see Fig. 6).

The temperature dependence of the exchange coupling is mainly governed by two competing mechanism. On the one hand, the effective exchange length increases with temperature, since the width $\delta_h$ of the hard FM domain wall increases [56]. This may result in inferior exchange coupling at low temperature ($\delta_h < t_s$), and rigid coupling at high temperature ($\delta_h \geq t_s$), due to the larger width of the domain wall [57, 58]. However, the extracted coercive field of the coupled Fe layer is smaller than the coercive field of the Dy layer for all considered temperatures. This suggests, that the bilayer is in the inferior coupling regime ($t_{Fe} > \delta_{Dy}$) also at higher temperatures, ruling out the described mechanism above.

On the other hand, the exchange coupling also depends on the anisotropy of the hard FM layer. If the Curie temperature of the hard FM layer is small, the anisotropy of the layer decreases with increasing temperature relatively fast. This leads, in contrast to the above described mechanism, to a degradation of the exchange coupling with increased temperature [59, 60].

Our experiments yield a decrease of the exchange coupled Fe layer's coercive field and critical field with increasing temperature. Since the Curie temperature of the employed Dy layer is small ($T_C \approx 160$ K), we can attribute the decrease of the exchange coupling to the degradation of the Dy layer's anisotropy. This assumption is also underlined by the temperature dependence, which we obtained for the Dy single layer (see. Fig. 3).

All in all, we could observe exchange coupling between the Fe and the Dy layer throughout the whole investigated temperature range. We also observed that the temperature can be used as a parameter to control the exchange coupling in the Fe/Dy bilayer. Since Dy is in the FM phase at $T$ = 120 K, the observed coupling is of FM/FM type for all temperatures.

### C. DEPENDENCE ON THE THICKNESS OF THE FE LAYER

For AFM/FM bilayers it was found that the strength of the exchange field $H_{ex}$ scales inversely $H_{ex} \propto (t_{FM})^{-1}$ with the thickness $t_{FM}$ of the FM layer (for $t_{AFM}$ = const.) [10, 12].

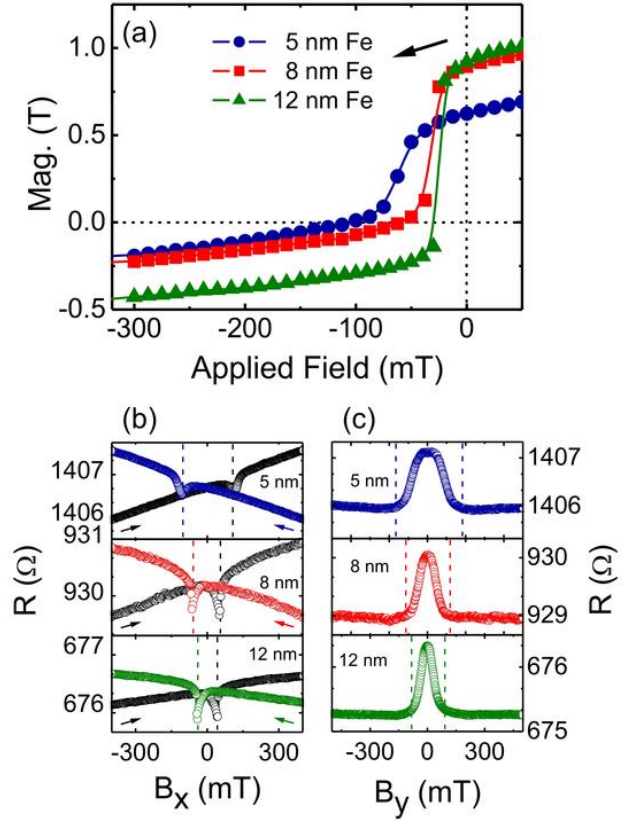

FIG. 9. (a) SQUID hysteresis loops (downsweep) of Fe/Dy full film samples with different thicknesses $t_{Fe}$ of the Fe layer (5 nm, 8 nm, 12 nm) and fixed $t_{Dy}$ (35 nm), obtained at 5 K. (b) Coercive field measurements at 4.2 K of single Fe/Dy stripes ($W$ = 3 μm) with the same layer sequence. Dashed lines mark the corresponding coercive field $B_{x,coerc}$. (c) Critical field measurements at 4.2 K for the same Fe/Dy stripes. Dashed lines mark the critical field $B_{x,crit}$.

A qualitatively similar scaling can also be observed for soft/hard FM bilayer systems. Here, the coercive field of the exchange coupled soft FM layer scales (for a given thickness of the hard FM layer) inversely with its thickness $t_s$ and is given by [13, 18]

$$H_{coerc} = \frac{\pi^2 A}{2\mu_o M_s} \cdot (t_s)^n \ (n < 0). \quad (2)$$

Equation (2) holds if $t_s$ is larger than the domain wall thickness $\delta_h$ of the hard FM layer. Here, $A$ is the exchange constant and $\mu_0 M_s$ is the saturation magnetization of the soft FM. For ideal systems, where the soft FM layer has no anisotropy and the hard FM is perfectly rigid, it was found that $n$ = -2 [13, 20]. Micromagnetic calculations for non-ideal systems, where the hard FM layer has a finite anisotropy, yield $n$ = -1.75 for thick soft FM layers ($t_s > 4\delta_h$) [54].

In the previous section, temperature was used to tune the exchange coupling. According to Eq. (2), the exchange coupling can also be controlled by the thickness $t_s$ of the employed soft FM layer. Therefore, we have grown and characterized Fe/Dy bilayers, where the thickness of the Dy layer was fixed at 35 nm and the thickness of the Fe layer $t_{Fe}$ was increased from 2.5 nm to 15 nm. This allows us to systematically study the dependence of the exchange coupling on $t_{Fe}$ in the Fe/Dy bilayer system. By evaluating $B_{coerc}$ ($t_{Fe}$) and $B_{crit}$ ($t_{Fe}$), we are able to determine the scaling of the coercive field with $t_{Fe}$.

Figure 9(a) shows SQUID hysteresis loops (downsweep) of Fe/Dy bilayers with different thicknesses of the Fe layer and constant thickness of the Dy layer ($t_{Dy}$ = 35 nm). The coercive field of the curves strongly depends on the Fe layer's thickness and decreases with $t_{Fe}$, as it is predicted by Eq. (2). The magnetoresistance measurements with microstructured FM stripes ($W$ = 3 μm) were carried out at $T$ = 4.2 K, and we applied the same methods as in the previous section.

The shape of the curves obtained from the coercive field measurements (not shown here) resembles for all bilayers those of the Dy single layer. The extracted Dy coercive field remains roughly constant for all $t_{Fe}$. This indicates that the Fe layer does not noticeably influence the magnetic properties of the hard FM Dy layer, even for larger thicknesses of the Fe layer. Figure 9(b) shows a zoom of the $B_x$ curves, which were obtained for different thicknesses of the Fe layer. All curves yield a relative minimum / dip in the magnetoresistance (marked with dashed lines). A similar feature / trace of the magnetoresistance curve has also been recorded for the bilayer discussed in the previous section [2.9 nm Fe / 35 nm Dy, see Fig. 7(c)], which allows to determine the coercive field of the coupled Fe layer. The extracted coercive fields (marked with dashed lines) decay with increasing thickness of the Fe layer, and are in good agreement for all thicknesses $t_{Fe}$ with the SQUID data.

Figure 9(c) shows magnetoresistance measurements of the critical field for different thicknesses of the Fe layer. Here, the magnetization of the bilayer was set along the stripe's axis, before a transverse field $B_y$ was applied. All curves exhibit a transition of the signal from a maximum to a relative minimum, as it has been observed for the bilayer discussed in the previous section [2.9 nm Fe / 35 nm Dy, see Fig. 7(d)]. The observed features can clearly be attributed to the critical field of the coupled Fe layer and are in agreement with the positive sign of the AMR effect of Fe. The critical field (marked with dashed lines) also decreases with the thickness of the Fe layer, as it has been observed for the coercive field measurements.

We also note, that the amplitude of the Fe AMR features becomes smaller for thinner Fe layers (especially for $t_{Fe} \leq$ 2.9 nm). For $t_{Fe}$ < 2.9 nm we are no longer able to determine the critical field of the coupled Fe layer, since the amplitude of the Fe AMR signal drops and gets superimposed by the Dy signal.

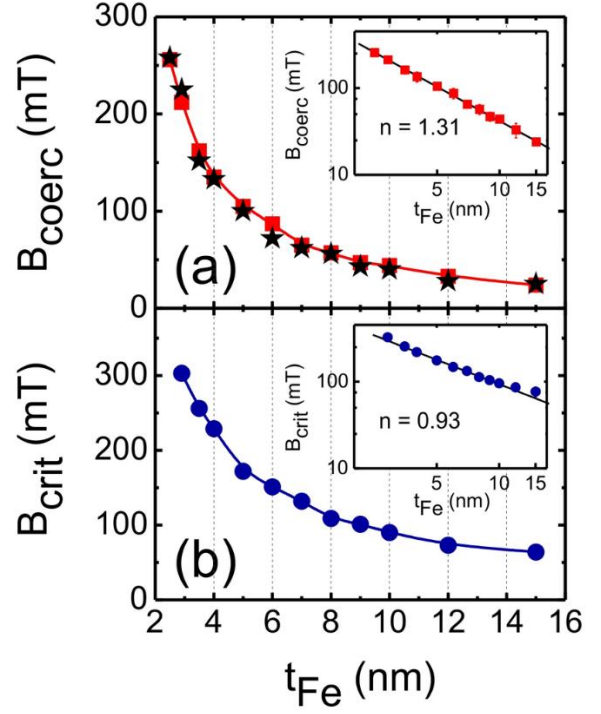

FIG. 10. (a) Coercive field $B_{x,coerc}$ ($t_{Fe}$) (squares) at 4.2 K for Fe/Dy bilayers with different thicknesses $t_{Fe}$ of the Fe layer and constant thickness of the Dy layer (35 nm). SQUID data of the coercive field (stars) is obtained from similarly grown full film samples. The line is a guide for the eye. (Inset) Log-log plot of $B_{x,coerc}$ ($t_{Fe}$). The exponent $n$ of the power law $B_{coerc}(t_{Fe}) \propto (t_{Fe})^n$ is deduced from the linear fit of the curve. (b) Critical field $B_{x,crit}$ ($t_{Fe}$) at 4.2 K for the same Fe/Dy bilayers as presented in (a). The line is a guide for the eye only. (Inset) Log-log plot of $B_{x,crit}$ ($t_{Fe}$). $n$ is deduced from the linear fit of the curve.

We can give two possible explanations for this observation. On the one hand, if the Fe layer thickness is comparable to the Dy domain wall width ($\delta_{Dy} \approx$ 2 nm [55]), both FM layers are fully coupled and reverse their magnetization at the same field. On the other hand, we can speculate that the AMR amplitude of the coupled Fe layer is too small to be observed for small $t_{Fe}$. According to Eq. (1), the ratio of the currents $I_{Fe}$ ($t_{Fe}$) and $I_{Dy}$ flowing through the Fe and Dy layer depends on the ratio of the individual layer resistances like $I_{Fe}$ ($t_{Fe}$) / $I_{Dy}$ = $R_{Dy}$ / $R_{Fe}$ ($t_{Fe}$). Here, the resistance $R_{Dy}$ of the Dy layer is fixed ($t_{Dy}$ = const.). If $t_{Fe}$ is reduced, the resistance $R_{Fe}$ of the Fe layer increases and the ratio $I_{Fe}$ ($t_{Fe}$) / $I_{Dy}$ decreases. Therefore, the current $I_{Fe}$ flowing through the Fe layer and the AMR amplitude decreases as well, when the Fe layer becomes thinner.

We now evaluate the dependence of the coupled Fe layer's coercive field on $t_{Fe}$. Fig. 10(a) shows the plot of the $B_{coerc}$ ($t_{Fe}$) curves, which include data obtained from SQUID (stars) and from magnetoresistance measurements (squares). Both curves show, that the coercive field decreases rapidly with increasing thickness of the Fe layer, as expected from Eq. (2). If plotted against $1/t_{Fe}$ the curve exhibits an almost linear dependence on $1/t_{Fe}$ (not shown). This suggests that the coupling is an interface effect, as observed for other soft/hard FM bilayer systems [61-63]. According to Eq. (2), the curves can be well described by a power law $B_{coerc}(t_{Fe}) \propto 1/(t_{Fe})^n$ $(n > 0)$.

The exponent $n$ is deduced from the log-log plot of $B_{x,coerc}(t_{Fe})$, shown in the inset of Fig. 10(a). We extract $n(B_{x,coerc}) = 1.31 \pm 0.01$ from the magnetoresistance data, and $n(B_{coerc, SQUID}) = 1.40 \pm 0.09$ from SQUID data. Both values are smaller than the value of $n$, which is theoretically predicted for non-ideal, exchange coupled systems ($n = -1.75$) [54]. Such deviations have also been reported for MBE grown materials [61]. In this study, the authors ascribe this mainly to the surface roughness at the interface, which affects the spin pinning.

The dependence of the critical field on $t_{Fe}$, shown in Fig. 10(b), can also by described with a power law $B_{crit}(t_{Fe}) \propto 1/(t_{Fe})^n$. The log-log plot of the critical field curve, from which we obtain $n(B_{x,crit}) = 0.93 \pm 0.02$, is shown in the inset of Fig. 10(b). We assume that the deviation between $n_{coerc}$ and $n_{crit}$ mainly stems from the fact, that both measurement routines are different in terms of magnetization reversal. To our knowledge, similar studies have not been reported yet in literature. However, the presence of linear $\ln(B_{crit})$ vs $\ln(t_{Fe})$-curves clearly indicates an exchange coupling mechanism [64].

All in all, the experiments presented in this section have shown that exchange coupling in the Fe/Dy bilayer can be controlled by changing the thickness of the Fe layer. This allows improving and engineering the coercive field and the critical field of the coupled Fe layer.

## VI. SUMMARY

We have performed a comprehensive study of the magnetic properties of sputtered Dy layers and their exchange coupling with thin, sputtered Fe layers at low temperatures (4.2 K - 120 K). Magnetoresistance and SQUID data prove that the deposited Dy single layer is of hard FM nature. Moreover magnetoresistance data exhibits a negative sign of the AMR effect of Dy, which has previously not been reported. We also observe a shift of the Curie temperature from 90 K to $T_C = 160$ K which is attributed to the growth conditions of the material. Measurements of a Fe/Dy bilayer (2.9 nm Fe / 35 nm Dy) yield an enhancement of the coercive field and of the critical field of the Fe layer by a factor of 14 - 22 (compared to the Fe single layer), which is due to exchange coupling between both layers. Data collected between 4.2 K and 120 K shows that the exchange coupling depends on the temperature, and persists even at 120 K, i.e. close to the Curie temperature of Dy. We also fabricated samples where the thickness of the Dy layer was fixed and the thickness of the Fe layer was varied between 2.5 nm and 15 nm. Here, data shows that the coercive field and the critical field of the coupled Fe layer scale inversely with its thickness, as predicted by theory. This allows engineering the coercive field and the critical field of the exchange hardened Fe layer by adjusting its thickness.

Overall, we have demonstrated that the coercive field and the critical field of a microstructured Fe layer can be enhanced and tailored, if brought in contact with a hard magnetic Dy layer. Thus, microstructured Fe/Dy bilayers could be widely employed in the field of spintronics, e.g. for SHE experiments, where precise control of the FM electrode's magnetic properties is necessary.

## ACKNOWLDEDGMENTS

We thank C. H. Back for fruitful discussions and valuable comments on the manuscript. This work has been supported by the Deutsche Forschungsgemeinschaft (DFG) via SFB 689.

---

* dieter.weiss@physik.uni-regensburg.de